\documentstyle[twoside, epsfig]{article}

\input ibvs2.sty
\begin{document}

\IBVShead{5955}{3 November 2010}

\IBVStitle{New Double-Mode and Other RR~Lyrae Stars from WASP Data}

\IBVSauth{Wils, Patrick$^1$}
\IBVSinst{Vereniging Voor Sterrenkunde, Belgium; e-mail: patrickwils@yahoo.com}

\SIMBADobj{V* UV Phe}
\SIMBADobj{HE 0414-2958}
\SIMBADobj{BPS BS 16478-0018}
\SIMBADobj{BPS BS 16466-0019}
\SIMBADobj{V* V633 Cen}
\SIMBADobj{V* V797 Her}
\SIMBADobjAlias{NSV 12753}{1SWASP J200447.74-371503.4}
\GCVSobj{V633 Cen}
\GCVSobj{V797 Her}
\GCVSobj{UV Phe}
\GCVSobj{NSV 12753}

\IBVStyp{RRd, RR(B)}
\IBVSkey{Variable stars, RR Lyrae stars, Double-mode variability}
\IBVSabs{42 RRab, 46 RRc and 7 previously unidentified double-mode RR~Lyrae stars were found in the publicly available data of the WASP archive.}
\IBVSabs{The Galactic double-mode RR~Lyrae stars appear to show a bimodal period distribution.}

\begintext

The data from the first public data release of the exoplanet transit survey WASP (Wide Angle Search for Planets; Butters et al., 2010)
were studied for a number of known and suspected RR~Lyrae stars (types RR, RRab and RRc), 
and for a number of Horizontal Branch stars (Beers et al., 1988, 1996, Christlieb et al., 2005),
in order to find previously unrecognized double-mode RR~Lyrae (RRd) stars.
In the analysis only TAMUZ (Collier Cameron et al., 2006) corrected data were used for which the uncertainty on the magnitude was less than 0.1.
The period analysis was done using {\sc Period04} (Lenz \& Breger, 2005).

Seven previously unidentified RRd stars were found in this way: four among previously known RR~Lyrae stars, 
and three more among the Horizontal Branch stars.  Details of these seven stars are all listed in Table~\ref{List}.  
After the name the WASP identification is given, followed by the full magnitude range (unfiltered WASP magnitude) and
the periods of the fundamental and first overtone modes.  
The last column contains a sequence number used in Tables~\ref{Amplitudes} and \ref{Phases}. 
These tables contain respectively the amplitudes and phases of the detected frequencies.  
Uncertainties are given between parentheses in units of the last decimal.  
These were calculated using the Monte Carlo simulations provided by {\sc Period04}.
As is usual for RRd stars, the first overtone mode has a larger amplitude than the fundamental mode,
except in the case of HE~0414-2958 = BPS~CS~22182-17, in which both frequencies have similar amplitudes.
As an illustration, phased light curves of V797~Her are provided in Fig.~1.
Note also that BPS~BS~16478-18 = BPS~BS~16553-34 has a double identifier in Beers et al. (1996).

Among the 3670 Horizontal Branch stars from Beers et al. (1988, 1996) and Christlieb et al. (2005) 
for which there were enough WASP observations,
108 RRab, 77 RRc and 5 RRd stars were identified.  Not all of these RR~Lyrae stars are however new discoveries.
Besides the three RRd stars from those catalogues mentioned in Table~\ref{List}, 
also the RRd stars BS~Com = BPS~BS~15626-36 (D\'ek\'any, 2007) 
and GSC~7509-299 = BPS~CS~22888-11 (Bernhard \& Wils, 2006) were recovered.
The relatively high number of RRc stars compared to the number of RRab stars may be a selection effect, 
as the objective prism and interference filter technique with which these stars were identified may favour the hotter RRc stars.
For comparison, the Large Magellanic Cloud contains 17693 RRab, 4958 RRc and 986 RRd stars (Soszy\'nski et al., 2009),
while the Small Magellanic Cloud contains 1933 RRab, 175 RRc and 258 RRd stars (Soszy\'nski et al., 2010).

\begin{table}
\begin{center}
\caption{New double-mode RR~Lyrae stars identified in WASP data.} \vskip2mm
\label{List}
\small
\begin{tabular}{lcrllr} 
\hline
Star & 1SWASP & \multicolumn{1}{c}{Range} & \multicolumn{1}{c}{$P_0$} & \multicolumn{1}{c}{$P_1$} & N \\
\hline
UV Phe & J$011210.74-411326.1$ & 14.3-14.8 & 0.534254(5) & 0.398668(1) & 1 \\
HE 0414-2958 & J$041649.58-295129.1$ & 14.4-15.0 & 0.477467(6) & 0.354824(4) & 2 \\
BPS BS 16478-18 & J$105743.63+384648.3$ & 13.6-14.4 & 0.494909(14) & 0.368498(5) & 3 \\
BPS BS 16466-19 & J$124322.85+345717.0$ & 13.9-14.6 & 0.486651(2) & 0.362228(1) & 4 \\
V633 Cen & J$141302.53-434817.7$ & 13.2-13.8 & 0.480517(3) & 0.357379(1) & 5 \\
V797 Her & J$171608.40+481752.7$ & 14.2-14.9 & 0.532299(5) & 0.397058(1) & 6 \\
NSV 12753 & J$200447.74-371503.4$ & 14.7-15.1 & 0.474658(11) & 0.353082(4) & 7 \\
\hline
\end{tabular}
\end{center}
\end{table}

\IBVSfig{15cm}{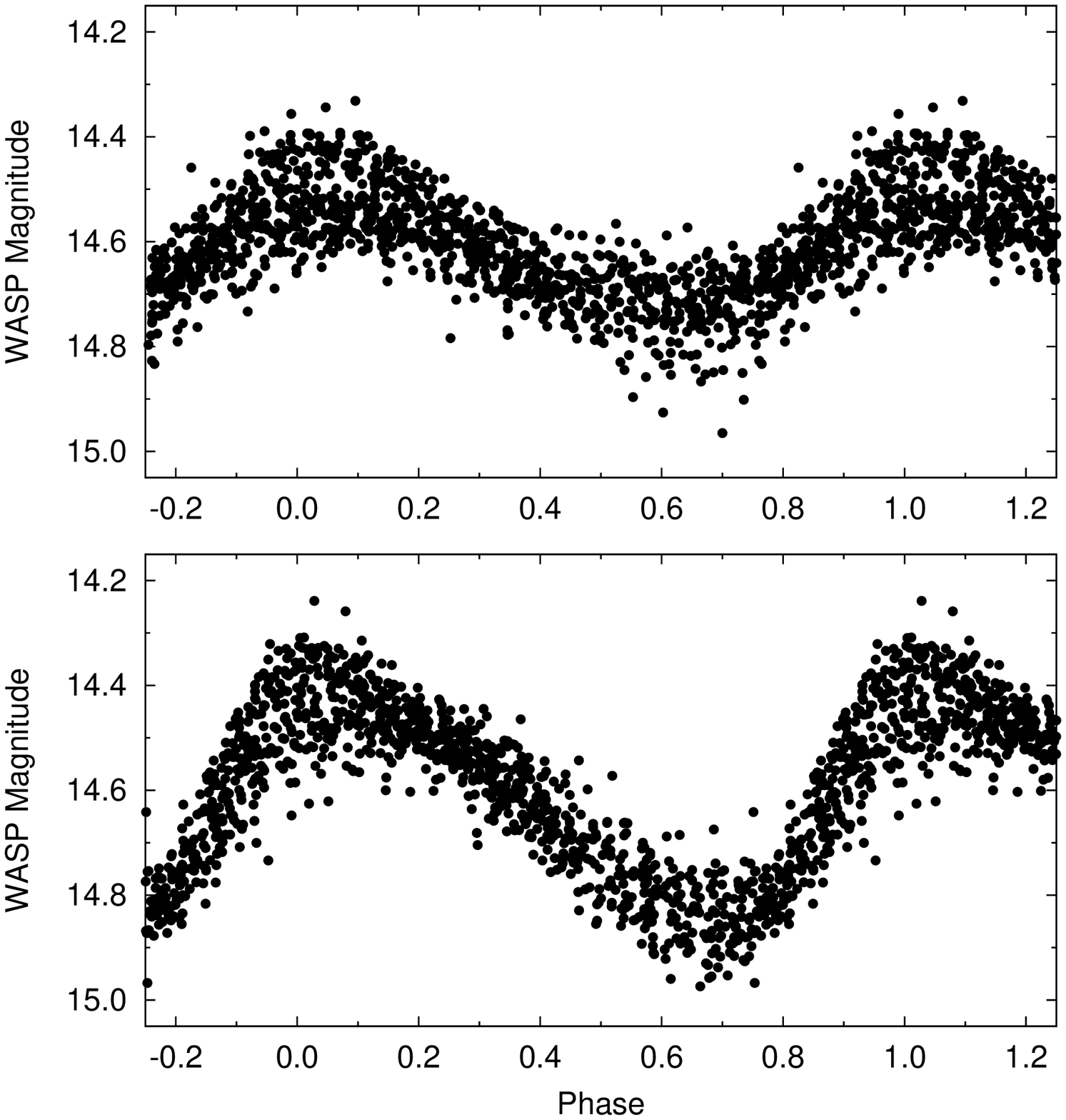}{Light curve of V797~Her. 
Top: phased with the fundamental period and prewhitened with the first overtone mode and its harmonics. 
Bottom: as above, but now phased with the first overtone period and prewhitened with the fundamental mode and its harmonics. 
Each point is the average of 10 consecutive WASP observations.}
\IBVSfigKey{V797Her.eps}{V797 Her}{Light curve}

\begin{table}
\begin{center}
\caption{Semi-amplitudes of the frequencies detected in the WASP data of the new RRd stars.
The number above each column refers to the stars in Table~\ref{List}.} \vskip2mm
\label{Amplitudes}
\small
\begin{tabular}{clllllll} 
\hline
Freq. & \multicolumn{1}{c}{1} & \multicolumn{1}{c}{2} & \multicolumn{1}{c}{3} & \multicolumn{1}{c}{4} & \multicolumn{1}{c}{5} & \multicolumn{1}{c}{6} & \multicolumn{1}{c}{7} \\
\hline
$f_0$ & 0.074(2) & 0.164(5) & 0.164(3) & 0.132(5) & 0.118(2) & 0.103(2) & 0.053(4) \\
$f_1$ & 0.140(2) & 0.158(5) & 0.209(3) & 0.188(4) & 0.156(2) & 0.200(2) & 0.088(4) \\
$f_0+f_1$ & 0.032(2) & 0.071(4) & 0.061(3) & 0.057(4) & 0.045(2) & 0.041(2) & 0.022(4) \\
$f_1-f_0$ & 0.028(2) & 0.048(5) & 0.046(3) & 0.041(5) & 0.030(2) & 0.035(2) & \multicolumn{1}{c}{-} \\
$2f_0$ & 0.010(2) & 0.047(5) & 0.027(3) & 0.026(4) & 0.023(2) & 0.016(2) & \multicolumn{1}{c}{-} \\
$2f_1$ & 0.027(2) & 0.034(4) & 0.044(3) & 0.028(5) & 0.025(2) & 0.044(2) & 0.013(3) \\
$3f_1$ & 0.009(2) & \multicolumn{1}{c}{-} & \multicolumn{1}{c}{-} & \multicolumn{1}{c}{-} & 0.009(2) & 0.014(2) & \multicolumn{1}{c}{-} \\
$f_0+2f_1$ & 0.015(2) & 0.025(5) & 0.028(3) & \multicolumn{1}{c}{-} & 0.018(2) & 0.015(2) & \multicolumn{1}{c}{-} \\
$2f_0+f_1$ & \multicolumn{1}{c}{-} & 0.025(5) & 0.016(3) & 0.020(4) & 0.010(2) & \multicolumn{1}{c}{-} & \multicolumn{1}{c}{-} \\
$2f_0+2f_1$ & \multicolumn{1}{c}{-} & \multicolumn{1}{c}{-} & 0.018(3) & \multicolumn{1}{c}{-} & 0.010(2) & \multicolumn{1}{c}{-} & \multicolumn{1}{c}{-} \\
$3f_0+f_1$ & \multicolumn{1}{c}{-} & \multicolumn{1}{c}{-} & \multicolumn{1}{c}{-} & \multicolumn{1}{c}{-} & 0.006(2) & \multicolumn{1}{c}{-} & \multicolumn{1}{c}{-} \\
\hline
\end{tabular}
\end{center}
\end{table}

For completeness, the other RR~Lyrae stars found among the Horizontal Branch stars that were not included in the AAVSO Variable Star Index \IBVSshortlink{http://www.aavso.org/vsx}{VSX}
at the time of writing, are listed in Tables~\ref{RRab} (42 RRab) and \ref{RRc} (46 RRc). 
When stars appear in two of the Horizontal Branch star lists,  
the designation of Christlieb et al. (2005) is given in the tables.
Due to the fairly low resolution of the WASP instruments, in some cases the magnitude range and the WASP coordinates may be affected by nearby stars.

\begin{table}
\begin{center}
\caption{Phases of the detected frequencies of the new RRd stars.  
These are given following the convention used by {\sc Period04} and with $T_0$= HJD 2450000.} \vskip2mm
\label{Phases}
\small
\begin{tabular}{clllllll} 
\hline
Freq. & \multicolumn{1}{c}{1} & \multicolumn{1}{c}{2} & \multicolumn{1}{c}{3} & \multicolumn{1}{c}{4} & \multicolumn{1}{c}{5} & \multicolumn{1}{c}{6} & \multicolumn{1}{c}{7} \\
\hline
$f_0$ & 0.077(4) & 0.944(4) & 0.032(3) & 0.358(5) & 0.393(3) & 0.059(4) & 0.916(9) \\
$f_1$ & 0.111(2) & 0.678(4) & 0.030(2) & 0.346(4) & 0.836(2) & 0.767(2) & 0.181(6) \\
$f_0+f_1$ & 0.570(8) & 0.002(12) & 0.443(8) & 0.106(12) & 0.615(6) & 0.229(9) & 0.464(27) \\
$f_1-f_0$ & 0.890(9) & 0.624(17) & 0.864(9) & 0.896(16) & 0.322(10) & 0.558(10) & \multicolumn{1}{c}{-} \\
$2f_0$ & 0.491(28) & 0.242(16) & 0.443(16) & 0.114(26) & 0.156(13) & 0.539(21) & \multicolumn{1}{c}{-} \\
$2f_1$ & 0.737(11) & 0.869(22) & 0.577(10) & 0.265(22) & 0.222(10) & 0.076(7) & 0.760(38) \\
$3f_1$ & 0.297(32) & \multicolumn{1}{c}{-} & \multicolumn{1}{c}{-} & \multicolumn{1}{c}{-} & 0.519(27) & 0.331(23) & \multicolumn{1}{c}{-} \\
$f_0+2f_1$ & 0.126(19) & 0.139(32) & 0.913(15) & \multicolumn{1}{c}{-} & 0.910(17) & 0.477(22) & \multicolumn{1}{c}{-} \\
$2f_0+f_1$ & \multicolumn{1}{c}{-} & 0.360(28) & 0.904(27) & 0.929(33) & 0.393(27) & \multicolumn{1}{c}{-} & \multicolumn{1}{c}{-} \\
$2f_0+2f_1$ & \multicolumn{1}{c}{-} & \multicolumn{1}{c}{-} & 0.440(22) & \multicolumn{1}{c}{-} & 0.772(31) & \multicolumn{1}{c}{-} & \multicolumn{1}{c}{-} \\
$3f_0+f_1$ & \multicolumn{1}{c}{-} & \multicolumn{1}{c}{-} & \multicolumn{1}{c}{-} & \multicolumn{1}{c}{-} & 0.379(48) & \multicolumn{1}{c}{-} & \multicolumn{1}{c}{-} \\
\hline
\end{tabular}
\end{center}
\end{table}

\begin{table}
\begin{center}
\caption{New RR~Lyrae stars pulsating in the fundamental mode (RRab) identified in WASP data 
among Field Horizontal Branch Stars (Beers et al., 1988, 1996, Christlieb et al., 2005).
The epoch of maximum is given as HJD - 2450000.  The letter B at the end of the line denotes stars that show the Blazhko effect.} \vskip2mm
\label{RRab}
\small
\begin{tabular}{lcrlll} 
\hline
Star & 1SWASP & \multicolumn{1}{c}{Range} & \multicolumn{1}{c}{Period} & \multicolumn{1}{c}{Epoch} \\
\hline
BPS CS 22876-029 & J$000157.75-364042.4$ & 13.5-14.0 & 0.63752 & 4001.56 & B \\
HE 0001-4300 & J$000400.87-424356.7$ & 14.3-14.8 & 0.51895 & 3880.61 &  \\
HE 0007-3416 & J$000944.08-335920.1$ & 13.9-15.0 & 0.60493 & 4270.57 &  \\
HE 0147-3030 & J$014926.72-301600.0$ & 14.0-15.2 & 0.57693 & 3981.48 &  \\
HE 0155-2108 & J$015744.41-205346.5$ & 14.7-15.0 & 0.35943 & 4379.60 &  \\
HE 0200-4322 & J$020236.92-430755.8$ & 14.6-15.0 & 0.67014 & 3999.57 &  \\
HE 0210-3735 & J$021237.64-372112.7$ & 13.6-13.8 & 0.50770 & 4050.38 &  \\
HE 0314-2836 & J$031616.36-282534.8$ & 14.2-14.5 & 0.59432 & 4090.31 &  \\
HE 0332-2129 & J$033419.05-211959.8$ & 14.4-15.2 & 0.54191 & 4353.56 & B \\
HE 0333-4650 & J$033452.98-464023.5$ & 13.9-14.8 & 0.48951 & 4484.32 & B \\
HE 0441-3136 & J$044255.90-313118.3$ & 14.8-16.3 & 0.34337 & 4412.44 &  \\
HE 0443-2513 & J$044505.91-250823.0$ & 15.0-15.3 & 0.32996 & 4110.35 &  \\
HE 0504-3113 & J$050606.02-310953.5$ & 14.2-14.7 & 0.56835 & 4136.41 &  \\
HE 0510-4101 & J$051207.76-405759.8$ & 14.5-15.1 & 0.68467 & 4029.44 &  \\
HE 0549-3927 & J$055104.65-392620.9$ & 14.8-15.4 & 0.56970 & 4522.28 &  \\
HE 1015-2201 & J$101812.69-221619.7$ & 14.2-15.0 & 0.56692 & 4522.48 &  \\
HE 1104-3222 & J$110703.96-323902.0$ & 14.0-14.4 & 0.56947 & 3862.05 &  \\
BPS BS 16545-047 & J$111329.97+353434.7$ & 13.1-13.3 & 0.45695 & 4140.71 &  \\
HE 1111-2927 & J$111352.38-294334.3$ & 14.8-14.9 & 0.47236 & 4155.51 &  \\
HE 1112-1950 & J$111451.40-200704.1$ & 15.0-15.4 & 0.30855 & 4564.38 &  \\
HE 1157-2813 & J$115953.34-282929.3$ & 13.6-13.9 & 0.52957 & 3898.24 &  \\
HE 1157-2519 & J$115958.37-253547.3$ & 14.5-14.7 & 0.51239 & 3890.22 &  \\
HE 1233-2316 & J$123636.50-233238.7$ & 14.9-15.4 & 0.60441 & 4586.24 &  \\
HE 1239-2151 & J$124201.81-220748.5$ & 12.4-12.5 & 0.55074 & 4495.54 &  \\
HE 1338-2727 & J$134100.84-274233.2$ & 14.2-15.1 & 0.62896 & 4562.38 &  \\
HE 1351-2348 & J$135421.23-240323.4$ & 14.0-14.8 & 0.46064 & 4588.43 &  \\
HE 1354-2320 & J$135702.57-233448.1$ & 14.6-15.1 & 0.60405 & 4562.50 &  \\
HE 1358-2125 & J$140049.44-214009.5$ & 13.9-15.1 & 0.47163 & 4572.53 &  \\
BPS BS 16554-067 & J$140655.96+205658.6$ & 13.6-13.7 & 0.33906 & 4261.54 &  \\
BPS CS 22936-279 & J$190129.86-354511.2$ & 13.8-14.2 & 0.64767 & 4250.67 &  \\
BPS CS 22885-084 & J$202501.21-422417.9$ & 14.4-14.6 & 0.49462 & 4387.26 &  \\
BPS CS 22955-119 & J$203041.85-235720.2$ & 14.1-14.7 & 0.60865 & 3960.46 & B \\
BPS CS 22955-139 & J$203637.65-240537.0$ & 13.7-15.1 & 0.46782 & 4301.35 &  \\
BPS CS 22880-004 & J$203911.98-212449.7$ & 13.6-14.4 & 0.61036 & 3891.58 &  \\
BPS CS 29501-046 & J$211204.06-370008.1$ & 14.0-14.8 & 0.54534 & 4271.70 &  \\
BPS CS 22948-023 & J$213530.97-390630.5$ & 14.5-15.0 & 0.57652 & 4300.53 & B \\
BPS CS 22948-084 & J$214600.39-401553.4$ & 14.0-15.0 & 0.66984 & 4300.39 &  \\
HE 2150-3053 & J$215320.37-303914.1$ & 14.1-14.7 & 0.64148 & 3960.32 &  \\
HE 2217-3717 & J$222042.00-370204.2$ & 15.1-15.2 & 0.55321 & 3999.29 &  \\
HE 2317-4517 & J$231959.89-450045.8$ & 14.1-14.3 & 0.62527 & 3954.56 &  \\
HE 2325-4624 & J$232746.99-460800.7$ & 14.4-15.0 & 0.29106 & 3909.61 & B \\
BPS CS 22876-023 & J$235742.03-340111.2$ & 14.4-15.2 & 0.27802 & 3953.60 &  \\
\hline
\end{tabular}
\end{center}
\end{table}

\begin{table}
\begin{center}
\caption{New RR~Lyrae stars pulsating in the first overtone mode (RRc).
For details, see Table~\ref{RRab}.} \vskip2mm
\label{RRc}
\small
\begin{tabular}{lcrlll} 
\hline
Star & 1SWASP & \multicolumn{1}{c}{Range} & \multicolumn{1}{c}{Period} & \multicolumn{1}{c}{Epoch} \\
\hline
BPS CS 29509-039 & J$005441.47-281354.6$ & 14.0-14.2 & 0.27286 & 4083.31 &  \\
HE 0055-3951 & J$005749.63-393531.4$ & 14.0-14.4 & 0.36111 & 4041.51 &  \\
HE 0145-2946 & J$014754.61-293131.2$ & 14.5-14.9 & 0.34140 & 4000.45 &  \\
HE 0222-2507 & J$022440.34-245403.6$ & 14.7-15.1 & 0.29095 & 3996.58 &  \\
HE 0250-3150 & J$025212.05-313827.5$ & 14.8-15.2 & 0.29625 & 4007.53 &  \\
HE 0311-2333 & J$031347.95-232239.6$ & 14.4-14.8 & 0.33963 & 4353.55 &  \\
HE 0351-3512 & J$035256.91-350327.9$ & 14.7-15.2 & 0.30276 & 4421.28 &  \\
HE 0428-3926 & J$042952.36-392004.6$ & 13.7-13.8 & 0.27715 & 4488.35 &  \\
HE 0442-3801 & J$044357.89-375609.2$ & 14.2-14.5 & 0.27606 & 4444.55 &  \\
HE 0505-3833b & J$050712.81-382956.0$ & 14.0-14.2 & 0.27391 & 4454.30 &  \\
BPS BS 16473-027 & J$084337.67+465824.3$ & 14.6-14.9 & 0.28122 & 4501.39 &  \\
BPS BS 16468-023 & J$090600.07+392758.6$ & 14.0-14.4 & 0.35638 & 4092.66 &  \\
BPS BS 16468-121 & J$092321.37+383836.6$ & 14.6-15.0 & 0.29065 & 4157.65 &  \\
BPS BS 16927-028 & J$093240.67+422108.4$ & 12.2-12.6 & 0.25399 & 4533.39 &  \\
HE 1046-2228 & J$104833.88-224414.8$ & 14.5-14.9 & 0.33384 & 4110.55 & B \\
BPS BS 16478-017 & J$105520.01+383039.0$ & 14.3-14.8 & 0.27416 & 4203.42 &  \\
BPS BS 16545-066 & J$112042.66+344712.6$ & 12.7-13.0 & 0.33970 & 4168.41 &  \\
HE 1122-2844 & J$112439.10-290048.2$ & 13.9-14.2 & 0.37957 & 4572.13 &  \\
BPS BS 16077-009 & J$113525.85+304318.1$ & 13.3-13.7 & 0.31215 & 4167.42 &  \\
HE 1222-2649 & J$122535.83-270549.7$ & 14.3-14.8 & 0.33595 & 4572.50 &  \\
HE 1228-2341 & J$123046.18-235743.6$ & 14.7-14.9 & 0.30872 & 4558.58 &  \\
BPS BS 16032-029 & J$124643.41+282809.8$ & 14.1-14.6 & 0.35752 & 4216.63 &  \\
HE 1302-2257 & J$130442.80-231336.6$ & 13.8-14.2 & 0.27356 & 4554.32 &  \\
BPS BS 16938-029 & J$130508.41+391533.1$ & 14.5-14.9 & 0.31621 & 3153.38 &  \\
BPS BS 16076-087 & J$130753.93+221007.1$ & 14.0-14.5 & 0.40112 & 4218.38 &  \\
BPS BS 15623-004 & J$141022.29+254433.1$ & 14.3-14.6 & 0.34471 & 4216.42 &  \\
BPS BS 16084-087 & J$161635.47+542258.4$ & 12.0-12.3 & 0.30627 & 4626.60 &  \\
BPS CS 22936-325 & J$190329.08-332433.8$ & 14.1-14.4 & 0.31945 & 3919.59 &  \\
BPS CS 22955-036 & J$202146.61-234129.5$ & 14.3-14.5 & 0.38810 & 4002.28 &  \\
BPS CS 22943-115 & J$202210.82-451849.2$ & 13.4-13.5 & 0.32715 & 3897.43 &  \\
BPS CS 22885-200 & J$202301.74-415448.5$ & 14.8-15.0 & 0.31697 & 4272.51 &  \\
BPS CS 22955-094 & J$203037.02-271349.3$ & 14.9-15.1 & 0.26593 & 4592.52 &  \\
BPS CS 22880-076 & J$204825.07-204635.4$ & 14.5-14.9 & 0.27237 & 4361.27 &  \\
BPS CS 29501-083 & J$211730.39-351757.9$ & 14.9-15.2 & 0.36968 & 4292.43 &  \\
HE 2115-4535 & J$211910.34-452233.7$ & 14.5-14.9 & 0.30140 & 3999.35 &  \\
HE 2126-4428 & J$213012.04-441520.4$ & 13.9-14.5 & 0.30029 & 4238.60 &  \\
BPS CS 29495-050 & J$214006.88-265319.3$ & 14.3-14.7 & 0.32103 & 4296.67 &  \\
BPS CS 29495-090 & J$214935.24-233018.7$ & 13.9-14.2 & 0.26758 & 3925.49 &  \\
BPS CS 22951-097 & J$215736.58-453236.0$ & 13.9-14.2 & 0.31660 & 4364.42 &  \\
HE 2201-2717 & J$220442.48-270233.3$ & 14.3-14.9 & 0.29918 & 3965.34 &  \\
HE 2309-3753 & J$231214.23-373719.5$ & 14.6-15.1 & 0.29260 & 4273.54 &  \\
HE 2316-3757 & J$231917.06-374047.9$ & 13.7-14.1 & 0.34475 & 4338.42 &  \\
BPS CS 29496-026 & J$234648.67-300028.7$ & 14.5-15.0 & 0.29756 & 3943.60 &  \\
HE 2344-2511 & J$234735.06-245507.8$ & 14.1-14.4 & 0.37096 & 4352.52 &  \\
HE 2349-4236 & J$235223.86-422000.8$ & 14.7-15.0 & 0.26362 & 3919.53 &  \\
HE 2356-4456 & J$235856.59-444014.6$ & 14.4-14.8 & 0.33155 & 3953.54 &  \\
\hline
\end{tabular}
\end{center}
\end{table}

A Petersen diagram of all known Galactic RRd stars is plotted in Fig.~2.
Apart from the stars from Table~\ref{List} and those listed in the references cited by D\'ek\'any (2009), the RRd stars given by
Wu et al. (2005), Gruberbauer et al. (2007), Pilecki \& Szczygie{\l} (2007), Szczygie{\l}  \& Fabrycky  (2007),
McClusky (2008), Sokolovsky et al. (2009) and Khruslov (2010) are included, 
75 in total.
In addition the brightest RRd stars with respectively $I<18.0$ and $I<18.5$ in the LMC
and SMC catalogues (Soszy\'nski et al., 2009 and 2010) were considered to be Galactic foreground objects.  
This gives 14 additional stars (7 from each of the LMC and SMC lists).

Three Galactic RRd stars have a higher period ratio than what can be expected from the other stars.  
The one with the highest ratio, [C2001c]~vd05f715 (Cseresnjes, 2001), has noisy data, so that one of the frequencies may well turn out to be spurious.
\footnote{In fact the MACHO data for this object (MACHO 122.23470.655; Allsman \& Axelrod, 2001) show it to be a contact binary with a period of 0.8309d,
twice the period of the first overtone mode given by Cseresnjes (2001).}
The limited number of data points for [IGF2000]~91 (Wu et al., 2005) may have resulted in inaccurate frequencies as well.
The most interesting of the outlier objects is likely OGLE~BUL-SC39~V1568 (Mizerski, 2003), as  
the OGLE II data (Udalski et al., 1997 and Szyma\'nski, 2005) for this object show additional frequencies very close to the main ones, 
an indication of a rapidly changing period.
In that case, the period ratio may not be reliable.  More observations of this object are highly recommended.

\IBVSfig{9cm}{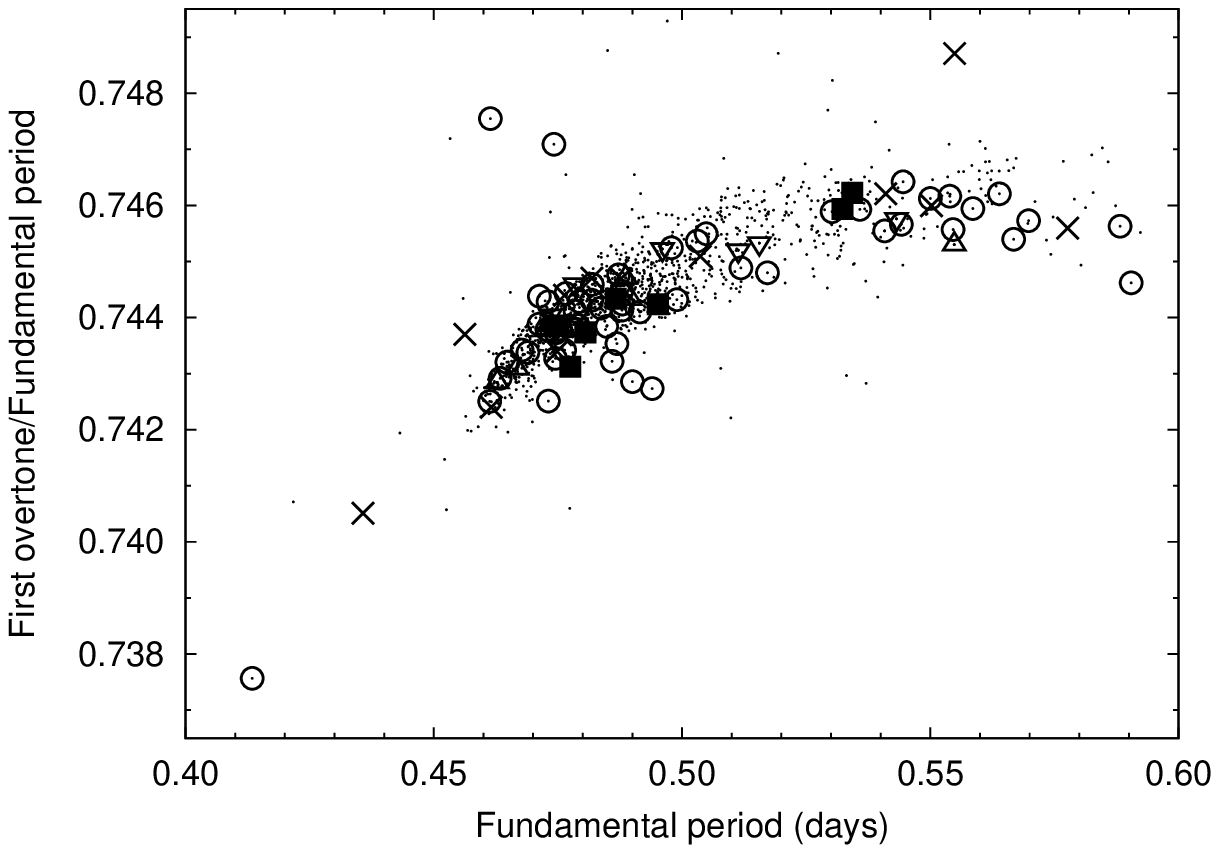}{Petersen diagram of 89 Galactic RRd stars.  
The stars from Table~\ref{List} are shown as black squares, foreground stars to the LMC and SMC as up- and downward pointing triangles resp., 
the Sgr foreground stars (Cseresnjes, 2001) as crosses, and other previously known RRd stars as open circles.
For comparison, the RRd stars of the LMC and SMC (Soszy\'nski et al., 2009, 2010) are plotted as small dots.}

The distribution of the periods of the Galactic RRd stars in Fig.~2 appears to be bimodal, 
with the majority of stars having a fundamental period around 0.48 days, a lack of stars with periods around 0.52-0.53 days
and a substantial number of stars with periods around 0.55 days.
This is clearly evident as well when comparing the cumulative distribution of the Galactic RRd stars 
with those from the Large and Small Magellanic clouds (Soszy\'nski et al., 2009, 2010) in Fig.~3. 
Although there is a small increase of SMC RRd stars with a period near 0.56 days, 
this increase is less pronounced than in the Galactic case.  
The LMC does not show this bimodality.

Because many of the Galactic RRd stars have been found in data from surveys that make at most a few observations per night, 
the apparent lack of RRd stars with a fundamental period near 0.52 days (or a first overtone period near 0.39 days) could be attributed to a selection effect.  
However it is more likely that variable stars with a period very close to an integer fraction of a day (e.g. 0.50 or 0.33 days) 
would go undetected.  
Also the RRab stars found in data from the Northern Sky Variability Survey (Wils et al., 2006 and Kinemuchi et al., 2006)
do not show fewer stars with periods near 0.39 or 0.52 days.
But the sample of Galactic RRd stars is certainly not a homogeneous sample as is the case for those found in the Magellanic Clouds, 
so this will need to be explored further.  
In addition, the relative number of known RRd stars with respect to RRab or RRc stars is still relatively small in the Galaxy 
compared to those in the Magellanic Clouds.

The RRd stars in the Sagittarius dwarf galaxy (Cseresnjes, 2001) show a similar bimodal distribution (Fig.~3).  
In this case the gap is symmetrically located around a fundamental period of 0.50 days, so that this could really be a selection effect as described above.
But again, the distribution of periods of RRab stars in the same field does not show a lack of stars with periods around 0.50 days (Cseresnjes et al., 2000).
Also the double-mode RR~Lyrae stars in the Sculptor dwarf galaxy (Kov\'acs, 2001) show a bimodal distribution, 
but with only 18 objects (including only two stars with a longer period) this sample is too small for definite conclusions.

The globular cluster IC~4449 contains only 13 short period RRd stars, 
while the 9 in M68 and the 8 in M15 have long periods (Clement et al., 1993).  
This could be explained in terms of the Oosterhoff dichotomy for those globular clusters (Oosterhoff 1939),
with longer period stars in metal-poor Oosterhoff type II clusters, and shorter period stars in relatively metal-rich Oosterhoff type I clusters.
The RRab population in the solar neighbourhood has been described as a mixture of metal-rich (Thick Disc), Oosterhoff I, and Oosterhoff II stars (Kinemuchi et al., 2006).
The bimodality in the period distribution of Galactic Field RRd stars may also be a consequence of this.

In Fig.~3 also an obvious shift of about 0.02 days can be seen between the average periods of the RRd stars in the SMC and the other galaxies (the LMC in particular).
The difference of the mean RRd period between the LMC and SMC is significant to better than the 99\% confidence level.  
It may be due to the different metallicities of stars in the Magellanic Clouds.
D\'ek\'any (2009) derived tight relations for the radius and density of double-mode RR~Lyrae stars 
as a function of their fundamental period.
Based on the longer period of the SMC stars, these relations indicate a higher mass, 
and from D\'ek\'any's (2009) mass-metallicity graphs also a lower metallicity on average for the SMC, 
as is generally accepted.

\IBVSfig{8.5cm}{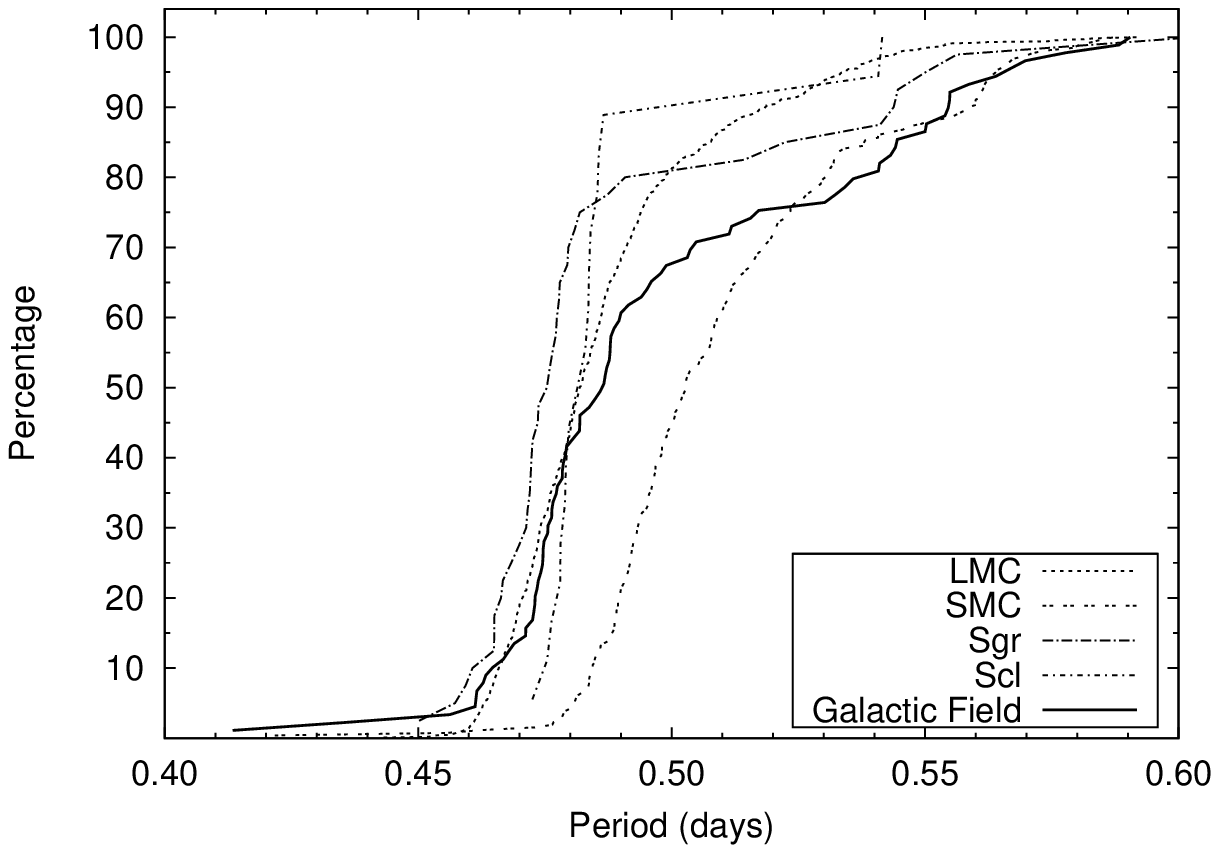}{Cumulative distribution of the periods of the fundamental mode of the double-mode RR~Lyrae stars 
in the Magellanic Clouds (Soszy\'nski et al., 2009, 2010), in the Sagittarius (Cseresnjes, 2001) and the Sculptor dwarf galaxies (Kov\'acs, 2001)
and in the Galactic Field.}

{\bf Acknowledgements:}
Data from the WASP public archive has been used in this research. 
The WASP consortium comprises of the University of Cambridge, Keele University, University of Leicester, The Open University, 
The Queen's University Belfast, St. Andrews University and the Isaac Newton Group. 
Funding for WASP comes from the consortium universities and from the UK's Science and Technology Facilities Council.

This study made use of data provided by the Simbad and VizieR databases maintained by the Centre de Donn\'ees astronomiques in Strasbourg (France).

The anonymous referee is acknowledged for useful suggestions to improve the paper.

\references

Allsman R.~A., Axelrod T.~S., 2001, arXiv:astro-ph/0108444v1 \BIBCODE{2001astro.ph..8444A}

Beers T.~C., Preston G.~W., Shectman S.~A., 1988, {\it ApJ Suppl.}, {\bf 67}, 461

Beers T.~C., Wilhelm R., Doinidis S.~P., Mattson C.~J., 1996, {\it ApJ Suppl.}, {\bf 103}, 433

Bernhard K., Wils P., 2006, {\it IBVS}, 5698

Butters O.~W., West R.~G., Anderson D.~R., et al., 2010, {\it A\&A}, {\bf 520}, L10

Christlieb N., Beers T.~C., Thom C., Wilhelm R., Rossi S., Flynn C., Wisotzki L., Reimers D., 2005, {\it A\&A}, {\bf 431}, 143


Clement C.~M., Ferance S., Simon N.~R., 1993, {\it ApJ}, {\bf 412}, 183

Collier Cameron A., Pollacco D., Street R.~A., et al., 2006, {\it MNRAS}, {\bf 373}, 799

Cseresnjes P., Alard C., Guibert J., 2000, {\it A\&A}, {\bf 357}, 871

Cseresnjes P., 2001, {\it A\&A}, {\bf 375}, 909

D\'ek\'any I., 2007, {\it Astron. Nachr.}, {\bf 328}, 833 \BIBCODE{2007AN....328..833D}

D\'ek\'any I., 2009, {\it AIP Conference Proceedings}, {\bf 1170}, 45 \BIBCODE{2009AIPC.1170..245D}

Gruberbauer M., Kolenberg K., Rowe J.~F., et al., 2007, {\it MNRAS}, {\bf 379}, 1498

Kinemuchi K., Smith H.~A., Wo\'zniak P.~R., McKay T.~A., 2006, {\it AJ}, {\bf 132}, 1202

Khruslov A.V., 2010, {\it PZP}, {\bf 10}, 11

Kov\'acs G., 2001, {\it A\&A}, {\bf 375}, 469

Lenz P., Breger M., 2005, {\it Comm. in Asteroseismology} {\bf 146}, 53 

McClusky J.V., 2008, {\it IBVS}, 5825

Mizerski T., 2003, {\it Acta Astron.}, {\bf 53}, 307

Oosterhoff P.~T., 1939, {\it The Observatory}, {\bf 62}, 104

Pilecki B., Szczygie{\l}  D.~M., 2007, {\it IBVS}, 5785

Sokolovsky K.V., Elenin L., Virnina N., 2009, {\it PZP}, {\bf 9}, 20

Soszy\'nski I., Udalski A., Szyma\'nski M.~K., et al., 2009, {\it Acta Astron.}, {\bf 59}, 15

Soszy\'nski I., Udalski A., Szyma\'nski M.~K., et al., 2010, {\it Acta Astron.}, {\bf 60}, 165

Szczygie{\l}  D.~M., Fabrycky  D.~C., 2007, {\it MNRAS}, {\bf 377}, 1263

Szyma\'nski M.~K., 2005, {\it Acta Astron.}, {\bf 55}, 43

Udalski A., Kubiak M., Szyma\'nski M., 1997, {\it Acta Astron.}, {\bf 47}, 319

Wils P., Lloyd C., Bernhard K., 2006, {\it MNRAS}, {\bf 368}, 1757

Wu C., Qiu Y.~L., Deng J.~S., Hu J.~Y., Zhao Y.~H., 2005, {\it AJ}, {\bf 130}, 1640

\endreferences

\end{document}